\documentclass[conference]{IEEEtran}
\IEEEoverridecommandlockouts

\usepackage{cite}
\usepackage{tabularx}
\usepackage{url}
\usepackage{amsmath,amssymb,amsfonts}
\usepackage{graphicx}
\usepackage{textcomp}
\usepackage{xcolor}
\def\BibTeX{{\rm B\kern-.05em{\sc i\kern-.025em b}\kern-.08em
    T\kern-.1667em\lower.7ex\hbox{E}\kern-.125emX}}

\begin{document}

\title{Probabilistic Analysis of Power Network Susceptibility to GICs
\thanks{This work was supported in part by an Open Philanthropy Project grant.}
}

\author{\IEEEauthorblockN{Michael Heyns\IEEEauthorrefmark{1}\IEEEauthorrefmark{2}, Stefan Lotz\IEEEauthorrefmark{2} and CT Gaunt\IEEEauthorrefmark{1}}
\IEEEauthorblockA{\IEEEauthorrefmark{1}Department of Electrical Engineering\\
University of Cape Town, Cape Town, South Africa}
\IEEEauthorblockA{\IEEEauthorrefmark{2}South African National Space Agency (SANSA), Hermanus, South Africa\\
mheyns@sansa.org.za}}

\maketitle

\begin{abstract}
As reliance on power networks has increased over the last century, the risk of damage from geomagnetically induced currents (GICs) has become a concern to utilities. The current state of the art in GIC modelling requires significant geophysical modelling and a theoretically derived network response, but has limited empirical validation. In this work, we introduce a probabilistic engineering step between the measured geomagnetic field and GICs, without needing data about the power system topology or the ground conductivity profiles. The resulting empirical ensembles are used to analyse the TVA network (south-eastern USA) in terms of peak and cumulative exposure to 5 moderate to intense geomagnetic storms. Multiple nodes are ranked according to susceptibility and the measured response of the total TVA network is further calibrated to existing extreme value models. The probabilistic engineering step presented can complement present approaches, being particularly useful for risk assessment of existing transformers and power systems.
\end{abstract}

\begin{IEEEkeywords}
Geomagnetically induced currents (GICs),\\empirical distributions, network risk modelling
\end{IEEEkeywords}

\section{Introduction}
Geomagnetically induced currents (GICs) in power networks are driven by variation in the geomagnetic field (\mbox{B-field}). This is the final link in a chain of coupled systems with their root in solar disturbances \cite{Albertson1974}. GIC modelling has two distinct steps, namely the geophysical step and the engineering step. The geophysical step aims to model the entire chain from Sun to ground conductivity and estimate the induced geoelectric field (E-field), which ultimately drives GICs. The engineering step uses the estimated E-field as input and models the network response, taking into account network specific factors such as topology and resistances, with transformer-level modelling being the state of the art \cite{Divett2018}. Recently, the geophysical step has been the subject of intense research in the space weather and geophysics communities, with strong focus on accurate E-field estimates based on detailed ground conductivity modelling \cite{Sun2019,Lucas2019}.
However, even with a very detailed and accurate E-field, the engineering step \cite{Gaunt2007,NERC} remains challenging as many factors regarding the network response are not known or are over simplified.

The approach we describe in this work does away with the two-step process, empirically linking concurrent B-field and GIC measurements, implicitly absorbing all driving factors without making the assumptions required by analytical GIC modelling \cite{Lehtinen1985}. Such analytical modelling has attempted to model the geophysical step as accurately as possible but does not make provision for probability distributions or uncertainty in parameters, particularly in the engineering step. Previous probability based analysis has been confined to B-field or \mbox{E-field} data, linked to GIC estimates purely through analytical modelling, and has focused mostly on extreme value analysis of possible GIC risk \cite{Thomson2011,Lucas2019,Pulkkinen2012} or hazard analysis \cite{Oughton2018}. Instead, a novel and practical method is used to analyse the susceptibility of a network to GICs by utilising (usually very limited) measured data sets in a way that provides probabilistic rather than exact estimates of the engineering step.

\begin{figure}[htbp]
\centerline{\includegraphics[width=3.4in]{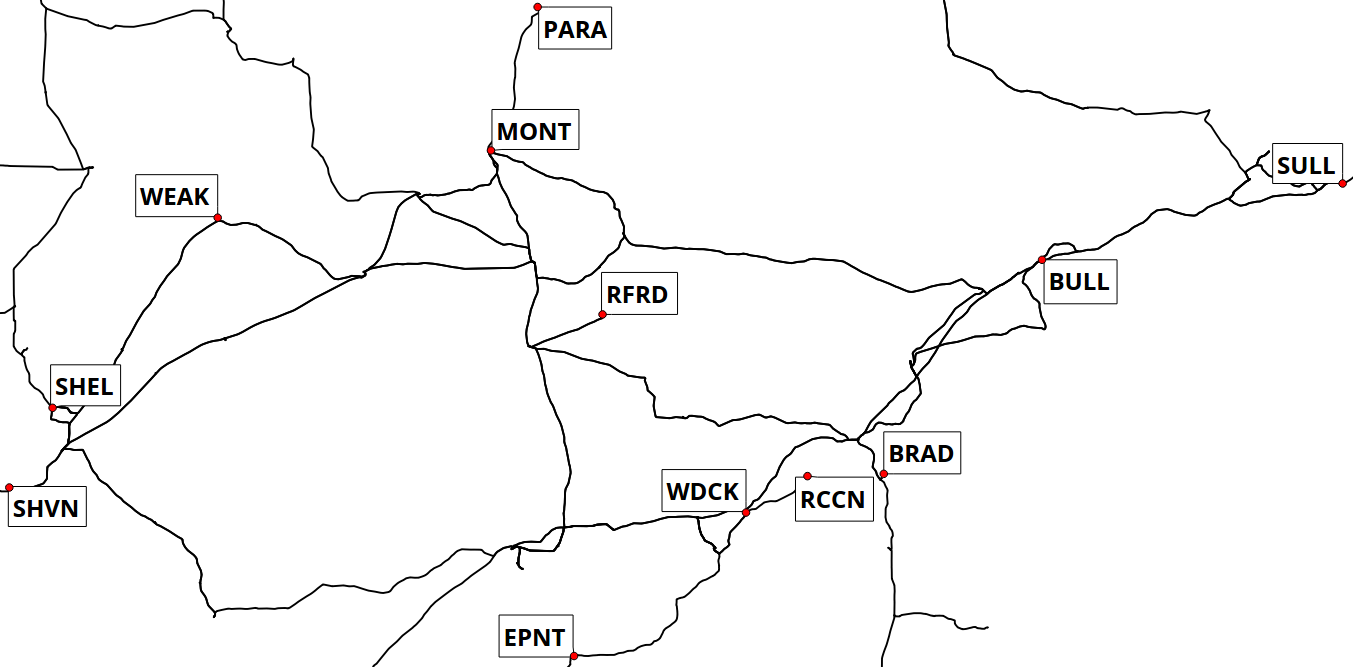}}
\caption{TVA HV network map with the substations analysed indicated.}
\label{fig:TVA}
\end{figure}
Measured GIC data used in this paper is from the Tennessee Valley Authority (TVA) network (see Figure \ref{fig:TVA}). Such typical mid-latitude networks are susceptible to GICs, as seen during the Halloween Storm of 2003 where there was limited damage in high-latitude regions but large accumulated transformer damage in mid-latitude southern African networks \cite{Gaunt2007}.

The rest of the paper is laid out as follows: Section \ref{sec:factor} describes the chain of events leading to GICs, and in Section \ref{sec:network} we lay out our novel approach to GIC modelling, which results in probabilistic estimates for network parameters. In Section \ref{sec:analysis}, the data used is described and results of analysis in the TVA network during 5 geomagnetic storms is presented. Finally, in Section \ref{sec:discuss} we discuss further implications of analysis in the TVA network and extrapolate exposure to extreme event scenarios.
\IEEEpubidadjcol

\section{Factor Chain Driving GICs}\label{sec:factor}
\begin{figure}[htbp]
\centerline{{\includegraphics[width=3.4in]{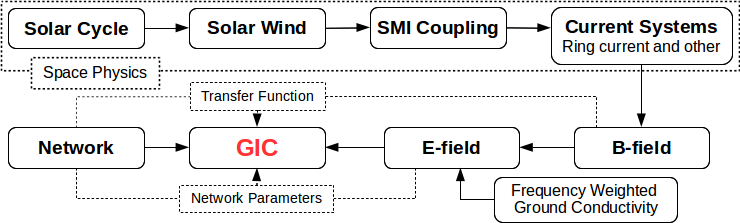}}}
\caption{A simplified factor flow for GICs in mid-latitude networks.}
\label{fig:flow}
\end{figure}
In Figure \ref{fig:flow}, the chain of events from solar activity to GIC is depicted, with emphasis on factors with large effects at mid-latitudes. Solar activity occurs in an 11-year {\em solar cycle}. During active periods, eruptions of plasma from the Sun known as coronal mass ejections (CMEs) expel plasma across space, impacting the near-Earth environment and causing geomagnetic storms. Geoeffectivity of plasma impacts are determined by the (i) position of the eruption on the solar disk, (ii) the conditions in the prevailing {\em solar wind}, and (iii) the {\em `solar wind-magnetosphere-ionosphere' (SMI) coupling}. If the near-Earth environment is still disturbed after a previous event, the effects from a follow-up event may be more intense than for otherwise quiet conditions. The {\em B-field} perturbations measured by mid-latitude observatories on the ground are due to the dynamics of near-Earth {\em current systems}. At mid-latitudes it is largely the east-west magnetospheric ring current that generates storm time perturbations \cite{deVilliers2017} for which the \mbox{SYM-H} (or lower resolution Dst) index is a coarse proxy. Ground perturbations in the {\em B-field} induce an {\em E-field}, modulated by the {\em frequency weighted ground conductivity} of the region \cite{Oyedokun2019}. The induced {\em E-field} then drives low-frequency {\em GICs}, which are ultimately dependent on the wider {\em network} characteristics, often defined by network parameters. An alternative method of modelling GICs is to use a transfer function between the B-field and GIC, absorbing ground conductivity and network parameter effects \cite{Ingham2017}.

In a probabilistic model of typical GIC exposure in a mid-latitude network, each link in the space physics chain could be assigned a probability distribution. For a coarse susceptibility estimate at mid-latitudes, the \mbox{SYM-H} distribution is representative of the driving geomagnetic storms. Cumulative SYM-H has further been linked to derived cumulative E-field activity in bulk extreme value studies \cite{Lotz2017}. In reality, further fine adjustments affect the local E-field and resulting GICs. Ground inhomogeneities and the coastal effect enhance the E-field along a geophysical strike. Local time plays a role, with the response to storm sudden commencement (SSC) and geomagnetic storm peak being different in different local time sectors. SYM-H however merges all these effects into a single proxy that characterises geomagnetic storms and can be calibrated against.

\section{Empirical Modelling of Network Factors}\label{sec:network}
After taking into account all the geophysical factors and deriving an E-field, the majority of current GIC modelling assumes a simplistic network model under dc driving \cite{Lehtinen1985}. Besides errors from the geophysical step propagating into this coarse approximation, the network also plays an active part in the GIC chain, with nodes influencing each other and even transformers influencing each other within nodes \cite{Divett2018}. Other factors include complex grounding, split driving in different transmission lines due to topology, quasi-ac driving and the general state of the system. Furthermore, there are medium-term temporal sensitivities as a network under stress from recent geomagnetic activity would have increased sensitivity to subsequent storms. This makes consecutive storms or moderate, but long duration events particularly dangerous.

GIC at a node at time $t$ can be modelled as, 
\begin{equation}
GIC(t)=\alpha E_x(t) + \beta E_y(t), \label{eq:gic}
\end{equation}
with the $\alpha$ and $\beta$ network parameters having units of [Akm/V] \cite{Lehtinen1985}. These network parameters scale the northward ($E_x$) and eastward ($E_y$) components of the \mbox{E-field} respectively, absorbing any errors in the geophysical modelling of the E-field and the network response. Assuming the E-field is perfectly aligned to the network, the absolute network parameter scaling would be $\sqrt{\alpha^2+\beta^2}$. Larger network parameters result in larger GICs for particular E-field components. A time series of simultaneous GIC and E-field measurements can be used to create an ensemble of $\alpha$ and $\beta$ estimates using pairwise combinations of linear equations represented by \eqref{eq:gic}. The resulting parameter ensembles define the effective network response, taking into account the entire network, non-trivially weighted. The TVA measurements, coupled with derived \mbox{E-field} data for the same region, provide a suitable dataset representative of a HV network under moderate GIC driving. 

\begin{figure}[htbp]
\centerline{\includegraphics[width=3.38in]{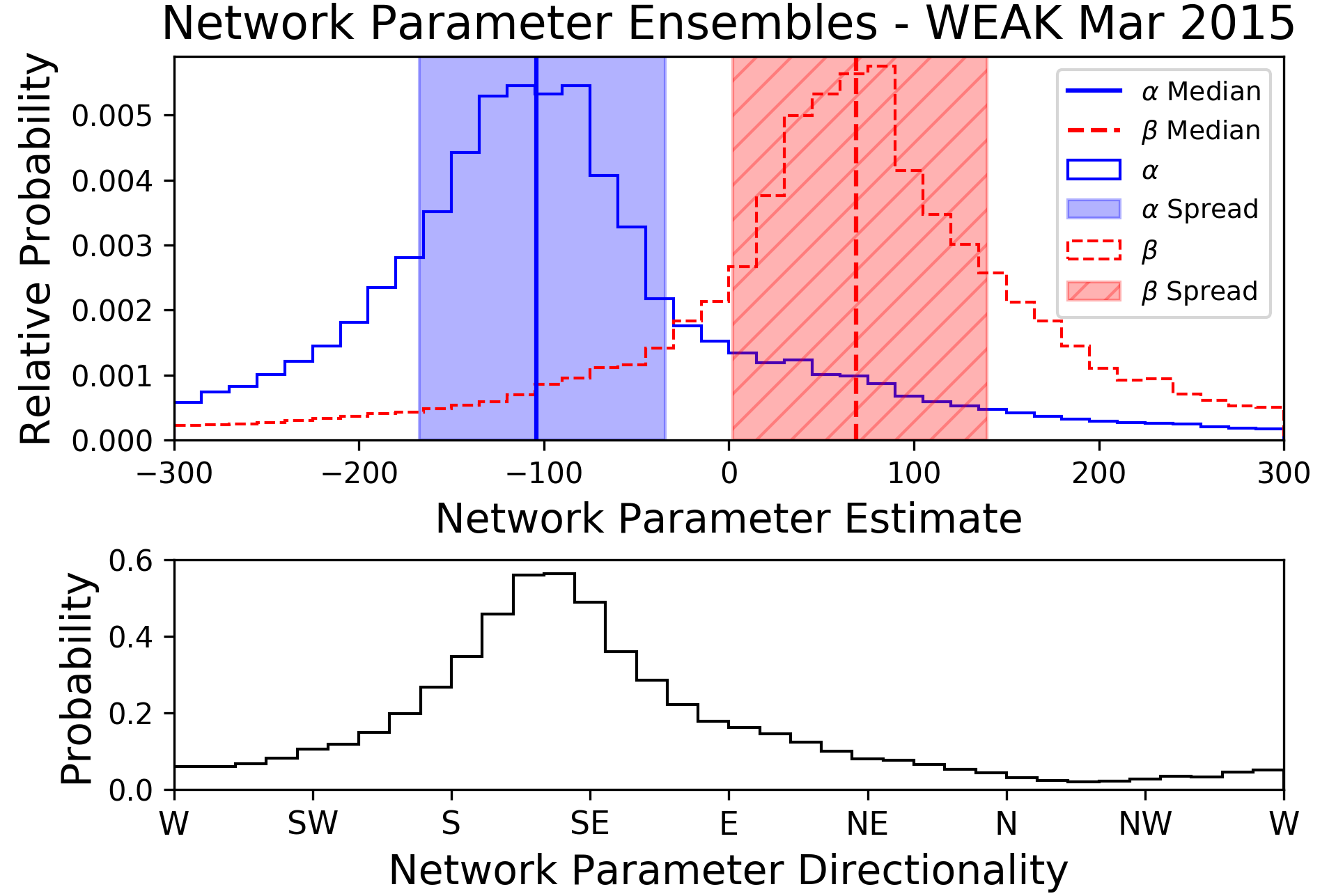}}
\caption{WEAK network parameter ensembles for the March 2015 geomagnetic storm, with the interquartile range defining the spread. Lower panel shows the effective network directionality with a local corner resulting in a SSE/SE peak.}
\label{fig:ens}
\end{figure}

This approach differs to previous modelling that assumes a single network parameter value. Coupled with the representative SYM-H distribution, we now have a coarse framework for susceptibility at a node. In this paper, for each event a random set of time instances above the median GIC level was chosen to produce an ensemble with a million estimates. Due to all combinations being used, only around \mbox{1 400} time instances are needed to produce the ensembles. Final estimates for nodal network parameters take the mean of multiple runs of different event ensembles, ensuring convergence. An example for the WEAK node in Mar 2015 is shown in Figure \ref{fig:ens}. For each ensemble, the most probable estimate is associated with the central peak, defined by the median of the distribution due to the distribution's heavy tails. The spread or variance in estimates is defined by the interquartile range and is driven by unmodelled aspects or errors in the modelling chain.

For each estimate of the network parameters, a further effective network directionality can be calculated,
\begin{equation}
\theta = \arctan{(\alpha/\beta)}. \label{eq:dir}
\end{equation}
The bearing $\theta$ takes into account the entire network and is the effective network direction that when aligned with the \mbox{E-field} creates the largest GICs, i.e. it modulates the driving E-field. No matter how large the \mbox{E-field}, if alignment is limited then so is the resulting GIC. An example of a directionality ensemble is shown for the WEAK node in Figure \ref{fig:ens}. Here two incident lines at a local corner contribute to the majority of the effective directionality, but the entire network is taken into account \cite{Overbye2013}. Since GICs are measured by a Hall-effect sensor on the transformer neutral, polarity is dependent of the set-up.

\section{Measured GIC Analysis in TVA Network}\label{sec:analysis}
GICs affect a network in two distinctly different ways. The impulsive effect from large peak GICs can result in thermal heating in transformers and possible voltage control maloperation. This effect has long been known and has been the topic of most GIC research and modelling, recently being the focus of the NERC benchmark for utility planning \cite{NERC}. A further effect not often taken into account is the cumulative damage from low-level driving, which can occur from GICs as low as 6 A \cite{Gaunt2007,Moodley2017}. Over an 11-year solar cycle, such accumulated damage is guaranteed -- ultimately resulting in accelerated ageing of transformers and premature failure. The state of the system, maintenance, age of existing equipment and previous GIC stress can all add to the impact of accumulated damage. In the scenario of a system operating above its capacity with excessive voltage control required, as is the case with loadshedding, susceptibility increases.

\subsection{Data}\label{subsec:data}
GIC data from substations in the TVA network have been used for large scale empirical validation. Such network-wide analysis differs from both full-network modelling, typically with no measured validation of the modelling, and the small scale validation of measurements at single nodes often done. The 2 s cadence GIC data have been cleaned for transient spikes and diurnal variation due to temperature. B-field data sampled at 1 s cadence from the nearest geomagnetic observatory (Fredericksburg) were resampled to 2 s cadence for consistency and used to derive the E-field using a global average conductivity profile \cite{Sun2015}. A global profile is not perfect but reproduces the relative frequency scaling expected when inhomogeneities in the ground conductivity are averaged over the induction footprint of the network \cite{Sun2019}, without any further modelling or measurement required. Such an approach is critical for utilities in regions with limited previous electromagnetic surveys. Any errors in the fine structure of the \mbox{E-field} should be consistent across the network with relative susceptibility still accurate, and more importantly comparable. 

\subsection{Geomagnetic Events}\label{subsubsec:event}
To determine the baseline GIC exposure in the TVA network, 5 different geomagnetic storms have been analysed. The first 3 are CME driven storms and are impulsive in nature, associated with peak GIC values. The last 2 events are co-rotating interaction region (CIR) driven storms, not often regarded due to their low-level of peak GIC activity but which may nevertheless lead to cumulative exposure. The characteristics of the events in terms of impulsive peak GIC exposure are summarised in Table \ref{tab:events}, and cumulative sustained GIC exposure in Table \ref{tab:events_cum}.
\begin{table}[htbp]
\caption{Geomagnetic Storms Characteristics}%
\begin{tabular}{m{\dimexpr.25\linewidth-2\tabcolsep-1.3333\arrayrulewidth}
|m{\dimexpr.16\linewidth-2\tabcolsep-1.3333\arrayrulewidth}
|m{\dimexpr.17\linewidth-2\tabcolsep-1.3333\arrayrulewidth}
|m{\dimexpr.16\linewidth-2\tabcolsep-1.3333\arrayrulewidth}
|m{\dimexpr.12\linewidth-2\tabcolsep-1.3333\arrayrulewidth}
|m{\dimexpr.13\linewidth-2\tabcolsep-1.3333\arrayrulewidth}}
\centering\textbf{Date\\(Type)} & \centering\textbf{SYM-H Min \scriptsize{[nT]}} & \centering\textbf{SYM-H Min \scriptsize{[UTC]}} & \centering\textbf{E-field Max \scriptsize{[mV/km]}}\\$E_x$ \textit{and}\\$E_y$ & \centering\textbf{GIC Max \scriptsize{[A]}} & \centering\textbf{TVA Node}  \arraybackslash \\
\hline
\hline
\rule{0pt}{2.3ex} \centering 11-13/09/2014 \\(CME) & \centering -97 & \rule{0pt}{2.3ex} \centering 23:03 12/09 & \rule{0pt}{2.3ex} \centering 93.48\\\textbf{131.56} & \centering 24.47 & PARA \arraybackslash \\
\hline
\rule{0pt}{2.3ex} \centering 16-22/03/2015 \\(CME) & \centering -234 & \rule{0pt}{2.3ex} \centering 22:47 17/03  & \rule{0pt}{2.3ex} \centering 89.80\\\textbf{113.97} & \centering 14.12 & MONT$^{\mathrm{a}}$ \arraybackslash \\
\hline
\rule{0pt}{2.3ex} \centering 22-29/06/2015 \\(CME) & \centering -208 & \rule{0pt}{2.3ex} \centering 04:24 23/06 & \rule{0pt}{2.3ex} \centering 74.16\\\textbf{168.14} & \centering 16.04 & PARA \arraybackslash \\
\hline
\rule{0pt}{2.3ex} \centering 05-09/10/2015 \\(CIR) & \centering -124 & \rule{0pt}{2.3ex} \centering 22:23 07/10 & \rule{0pt}{2.3ex} \centering \textbf{44.66}\\37.88 & \centering 9.19 & PARA \arraybackslash \\
\hline
\rule{0pt}{2.3ex} \centering 15-18/02/2016 \\(CIR) & \centering -60 & \rule{0pt}{2.3ex} \centering 00:28 18/02 & \rule{0pt}{2.3ex} \centering \textbf{36.08}\\35.14 & \centering 8.03 & PARA \arraybackslash \\
\hline
\multicolumn{6}{r}{\scriptsize{$^{\mathrm{a}}$No PARA data for given event}}
\end{tabular}
\label{tab:events}
\end{table}
\begin{table}[htbp]
\caption{Geomagnetic Storms Cumulative Characteristics}%
\begin{tabular}{m{\dimexpr.25\linewidth-2\tabcolsep-1.3333\arrayrulewidth}
|m{\dimexpr.14\linewidth-2\tabcolsep-1.3333\arrayrulewidth}
|m{\dimexpr.16\linewidth-2\tabcolsep-1.3333\arrayrulewidth}
|m{\dimexpr.19\linewidth-2\tabcolsep-1.3333\arrayrulewidth}
|m{\dimexpr.12\linewidth-2\tabcolsep-1.3333\arrayrulewidth}
|m{\dimexpr.14\linewidth-2\tabcolsep-1.3333\arrayrulewidth}}
\centering\textbf{Date  and \\Duration\\ \scriptsize{[Hours]}} & \centering\textbf{SSC Onset \scriptsize{[UTC]}} & \centering\textbf{SYM-H RMS\\\scriptsize{[nT]}} & \centering\textbf{E-field RMS \scriptsize{[mV/km]}}\\\scriptsize{$E_x$ \textit{and}}\\\scriptsize{$E_y$} & \centering\textbf{GIC\\RMS\\\scriptsize{[A]}} & \centering\textbf{TVA Node}  \arraybackslash \\
\hline
\hline
\rule{0pt}{2.3ex} \centering 11-13/09/2014 \\41.7 & \rule{0pt}{2.3ex} \centering 15:53\\12/09 & \centering 25.52 & \rule{0pt}{2.3ex} \centering 3.82\\\textbf{5.73} & \centering 0.77 & PARA  \arraybackslash \\
\hline
\rule{0pt}{2.3ex} \centering 16-22/03/2015 \\138.6 & \rule{0pt}{2.3ex} \centering 04:45
17/03 & \centering 71.65 & \rule{0pt}{2.3ex} \centering 5.54\\\textbf{6.03} & \centering 1.34 & MONT$^{\mathrm{a}}$ \arraybackslash \\
\hline
\rule{0pt}{2.3ex} \centering 22-29/06/2015 \\ 166.1 & \rule{0pt}{2.3ex} \centering 18:33
22/06 & \centering   68.38 & \rule{0pt}{2.3ex} \centering 4.83\\\textbf{6.03} & \centering 1.83  & PARA \arraybackslash \\
\hline
\rule{0pt}{2.3ex} \centering 05-09/10/2015 \\83.5 & \centering N/A$^{\mathrm{b}}$ & \centering 51.29 & \rule{0pt}{2.3ex} \centering 4.06\\\textbf{4.34} & \centering 1.55 & PARA \arraybackslash \\
\hline
\rule{0pt}{2.3ex} \centering 15-18/02/2016 \\ 69 & \centering N/A$^{\mathrm{b}}$ & \centering 35.20 & \rule{0pt}{2.3ex} \centering \textbf{3.76}\\3.23 & \centering 0.75 & PARA \arraybackslash \\
\hline
\multicolumn{6}{r}{\scriptsize{$^{\mathrm{a}}$No PARA data for given event}}\\
\multicolumn{6}{r}{\scriptsize{$^{\mathrm{b}}$CIR event with no obvious sudden storm commencement (SSC)}}
\end{tabular}
\label{tab:events_cum}
\end{table}

The span of each storm is defined as the period from sudden impulse, when an interplanetary shock hits the magnetosphere, through to when the magnetosphere recovers to quiet time levels, i.e. when SYM-H recovers to greater than -20 nT after having reached a minimum value at the peak of the storm \cite{Lotz2017}. The cumulative value of SYM-H is taken as the minutely RMS of the storm to allow for comparisons between storms of different lengths. Multiplying the RMS by the duration gives an idea of the total exposure for a single storm. To avoid noise levels, the GIC and E-field RMS values are defined as the 2 s cadence RMS above the median level for each.

\begin{figure}[htbp]
\centerline{\includegraphics[width=3.38in]{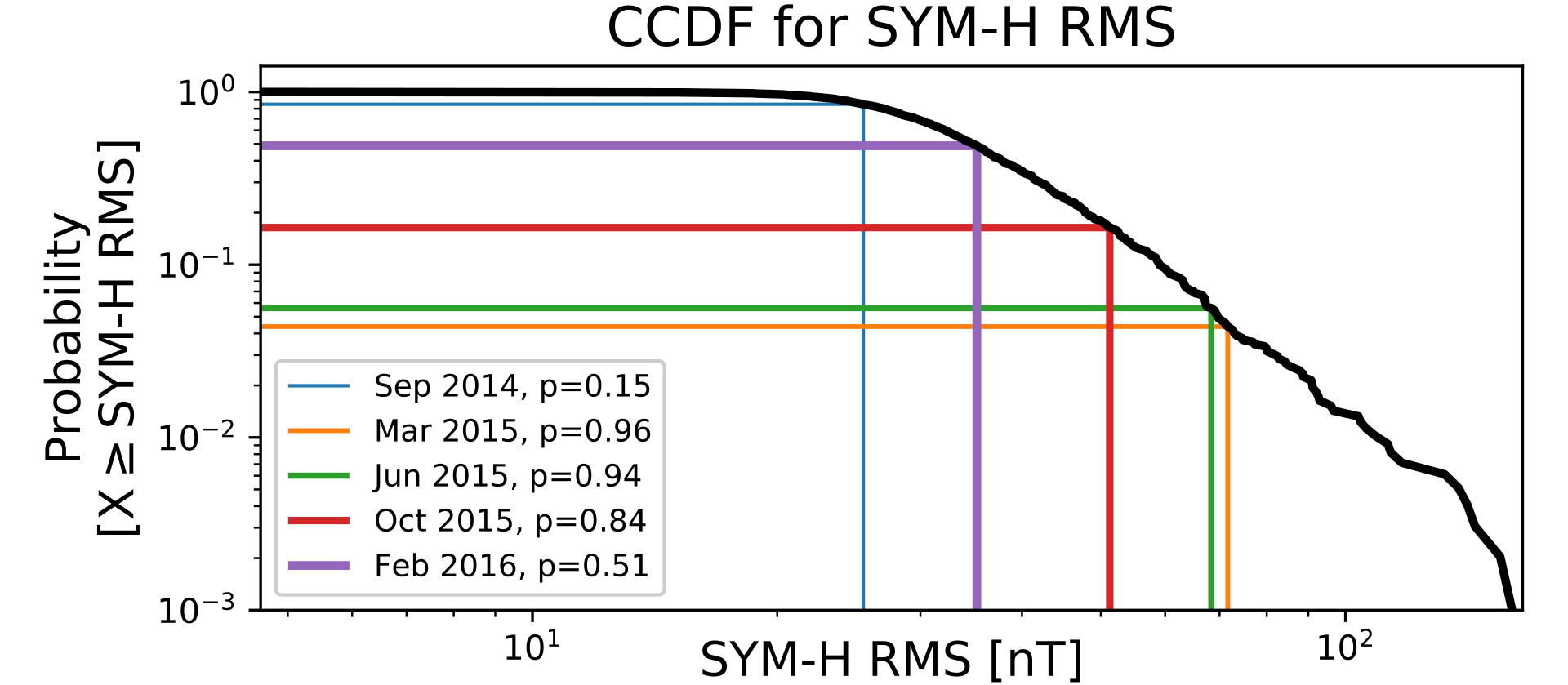}}
\caption{Complimentary cumulative distribution function (CCDF) of SYM-H RMS for all storms with minimum SYM-H$<-50$ nT for 3.5 solar cycles.}
\label{fig:symh_rms}
\end{figure}

To contextualise the 5 geomagnetic events, a complimentary cumulative distribution function (CCDF) is defined using all geomagnetic storms with minimum SYM-H $<-50$ nT between 1981 and 2018, identified according to the algorithm described by \cite{Lotz2017}. Figure \ref{fig:symh_rms} shows the CCDF of SYM-H RMS. The 5 events analysed are indicated with vertical lines and their probabilities are listed in the legend. The probability associated with each event indicates the fraction of events (totalling 981 in the 38 year interval) with RMS SYM-H smaller than the event. For example, 0.15 (15\%) of all events will be less intense than the weak cumulative Sep 2014 event, i.e. we expect about 0.85 or 241 events to be larger over the course of an average solar cycle, modulated by the peak occurrence at solar maximum and declining phase \cite{Echer2011}. For the most intense cumulative event, Jun 2015, roughly 11 larger events can be expected over a solar cycle. For the weakest event in terms of minimum \mbox{SYM-H} reached, Feb 2016, we expect around 210 larger events to occur per solar cycle. For Mar 2015, the most intense impulsive event at -234 nT, we expect roughly 10 more intense events per solar cycle.

\subsection{Measured Results and Risks}\label{subsec:measured}
\begin{figure}[htbp]
\centerline{\includegraphics[width=3.38in]{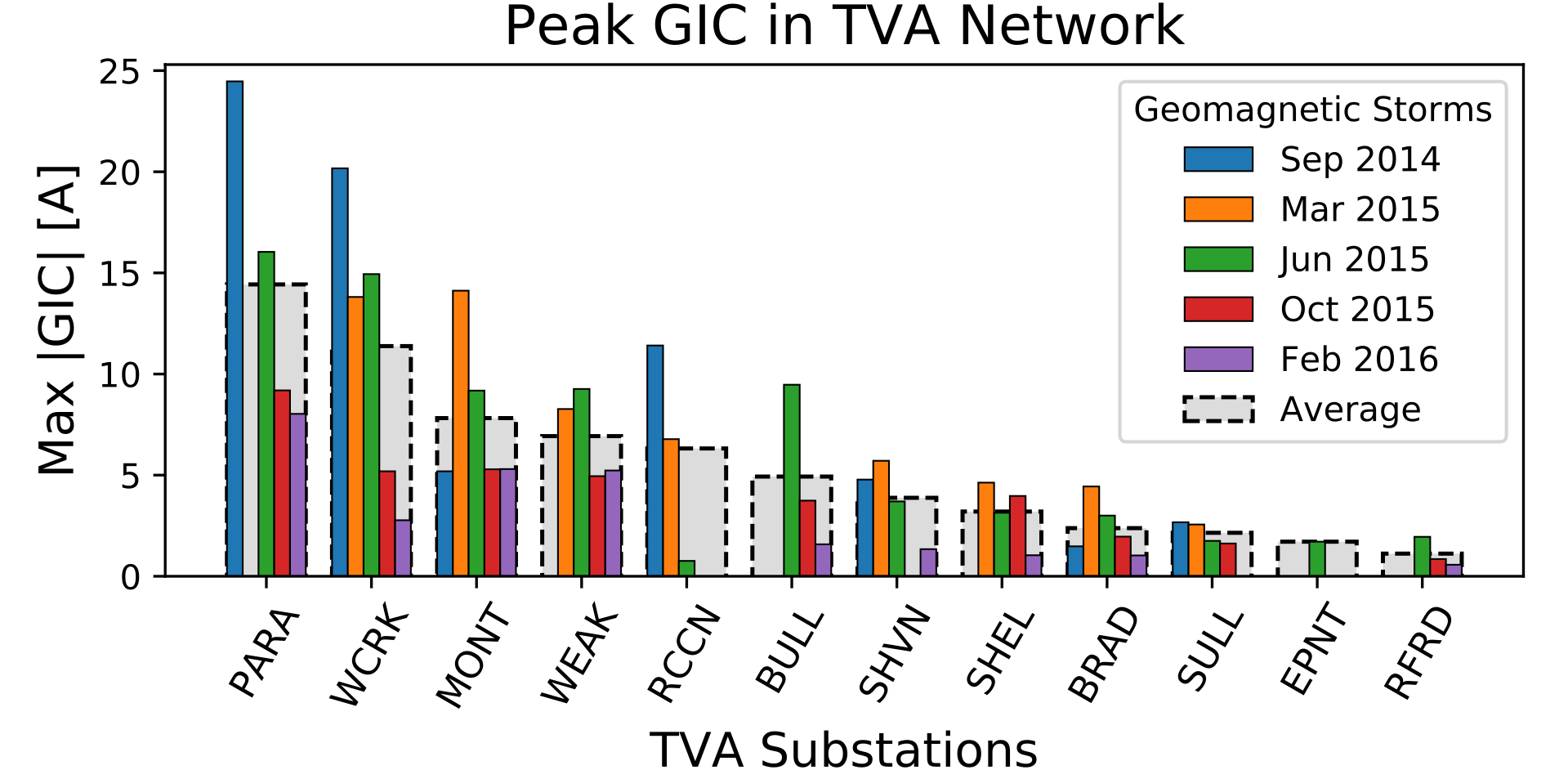}}
\caption{Peak GIC measured in the TVA network for 5 geomagnetic storms.}
\label{fig:max_stns}
\end{figure}

Using measured GIC data, the susceptibility of the various nodes in the TVA network can be ranked in terms of impulsive and cumulative exposure. In Figure \ref{fig:max_stns}, the maximum measured GIC at each node is shown per storm, with the CME storms having larger peak GICs. PARA, a terminal north-south node, is the most susceptible, having the entire network southward act as a catchment area. Other local corner nodes such as WEAK and WCRK are also more affected. Adjacent nodes can also be associated with the larger GIC flows at terminal nodes, as seen at MONT and RCCN. When a series capacitor is present, a line is effectively removed. Network information is needed to confirm such cases in the TVA network. 

\begin{figure}[htbp]
\centerline{\includegraphics[width=3.38in]{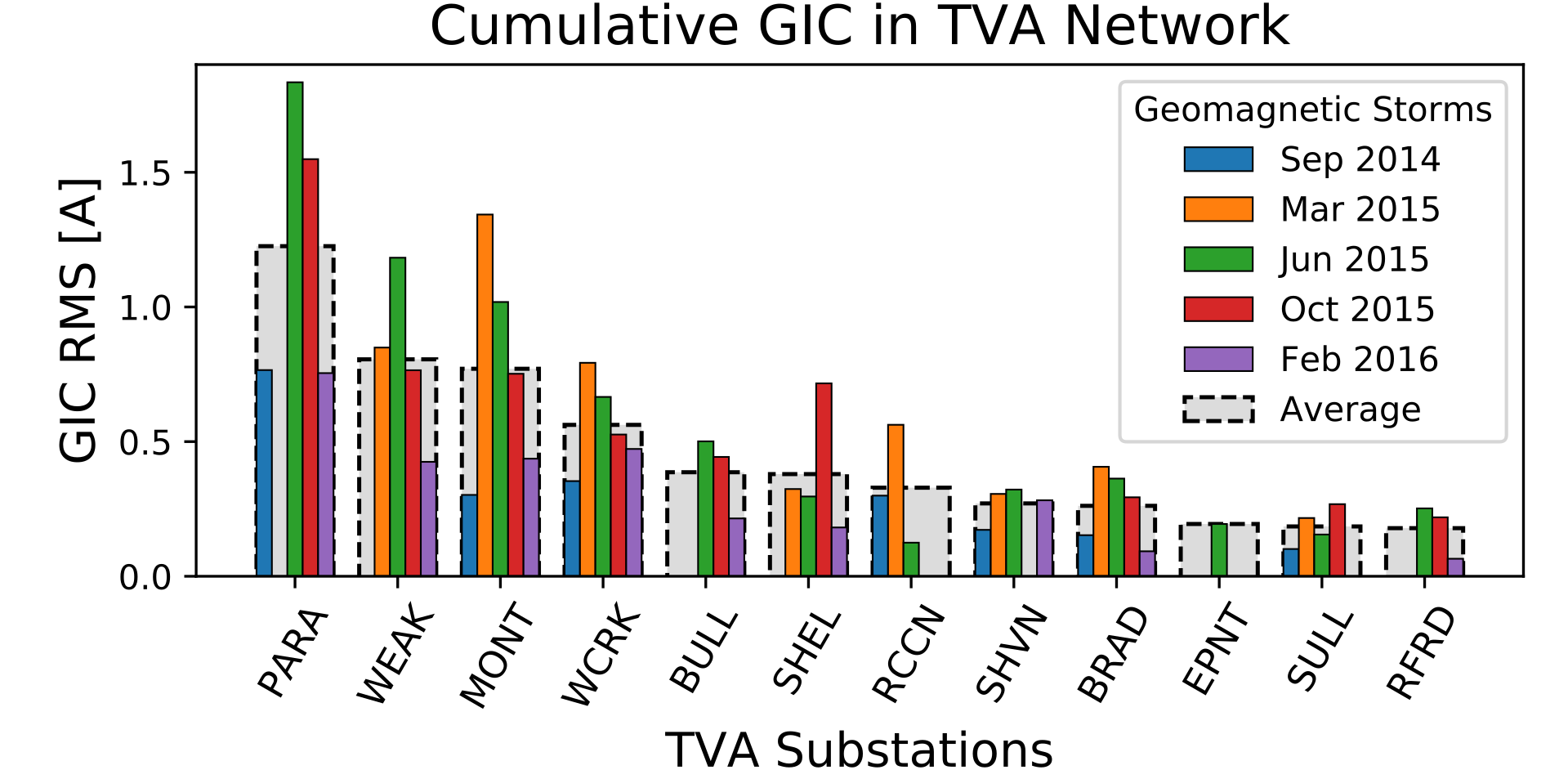}}
\caption{Cumulative GIC exposure in the TVA network, as defined by the RMS of GIC above the median GIC level for each geomagnetic storm.}
\label{fig:cul_stns}
\end{figure}

The cumulative exposure seen in Figure \ref{fig:cul_stns}, similarly shows PARA as the most susceptible node. Of interest is the difference in storm response, where the Sep 2014 storm is highly impulsive with large a peak GIC, the Jun 2015 storm has a larger cumulative effect. This may be due partly to a period of sustained long-period pulsation driving. Over all events, there is no clear or consistent pattern, suggesting the local network and the finer structure of a geomagnetic storm need to be taken into account. SYM-H identifies geomagnetic storms well, but no two storms are the same.

\subsection{Ensemble Modelled Results and Risk Ranking}\label{subsec:ens}
In order to characterise the network in finer detail, the network parameter ensembles defined in Section \ref{sec:network} are used. In order to minimise low-level noise at certain substations, the GIC and E-field data used for ensemble estimation (and trend fitting later in Figure \ref{fig:max_rel}) were resampled to 4 s cadence and only data above the median level used. The resulting network parameter scaling and effective directionality for each node given the event coverage is summarised in Table \ref{tab:results}.
\begin{table}[htbp]
\caption{Node Specific Ensemble Results and Risk in the TVA Network}%
\begin{tabular}{m{\dimexpr.17\linewidth-2\tabcolsep-1.3333\arrayrulewidth}
|m{\dimexpr.14\linewidth-2\tabcolsep-1.3333\arrayrulewidth}
|m{\dimexpr.14\linewidth-2\tabcolsep-1.3333\arrayrulewidth}
|m{\dimexpr.08\linewidth-2\tabcolsep-1.3333\arrayrulewidth}
|m{\dimexpr.17\linewidth-2\tabcolsep-1.3333\arrayrulewidth}
m{\dimexpr.17\linewidth-2\tabcolsep-1.3333\arrayrulewidth}
|m{\dimexpr.13\linewidth-2\tabcolsep-1.3333\arrayrulewidth}}
\rule{0pt}{2.3ex} \centering\textbf{Node$^{\mathrm{a}}$}\\(Risk & \centering\textbf{Geog.\\Lat.} & \centering\textbf{Geog.\\Lon.} & \centering \textbf{\textit{n}}$^{\mathrm{b}}$ &  \multicolumn{2}{m{\dimexpr.34\linewidth-2\tabcolsep-1.3333\arrayrulewidth}|}{\centering \textbf{Network Parameters} \\\textit{Median $\scriptstyle\pm$ Spread}} &  \centering \textbf{Bearing} \arraybackslash \\
\rule{0pt}{-0.5ex} \centering Rank) & & & & \centering $\boldsymbol\alpha$ & \centering $\boldsymbol\beta$ & \centering $\boldsymbol\theta$ \arraybackslash \\
\hline
\hline
\rule{0pt}{2.3ex} \centering \textbf{PARA} \\ (1.00) & \centering 37.3$^\circ$& \centering -87.0$^\circ$ & \centering 4 & \rule{0pt}{2.3ex} \centering -241.05 \\ $\scriptscriptstyle\pm$ 254.66 & \rule{0pt}{2.3ex} \centering -8.72 \\ $\scriptscriptstyle\pm$ 383.88 & \centering 182$^\circ$ \arraybackslash \\
\hline
\rule{0pt}{2.3ex} \centering \textbf{WEAK} \\ (0.50) & \centering 36.3$^\circ$ & \centering -88.8$^\circ$ & \centering 4 & \rule{0pt}{2.3ex} \centering -94.69 \\ $\scriptscriptstyle\pm$ 117.44 & \rule{0pt}{2.3ex} \centering 73.27 \\$\scriptscriptstyle\pm$ 146.45 & \centering 142$^\circ$ \arraybackslash \\
\hline
\rule{0pt}{2.3ex} \centering \textbf{MONT} \\ (0.48) & \centering 36.6$^\circ$ & \centering -87.2$^\circ$ & \centering 5 & \rule{0pt}{2.3ex} \centering -111.14 \\ $\scriptscriptstyle\pm$ 118.58 & \rule{0pt}{2.3ex} \centering 35.20 \\ $\scriptscriptstyle\pm$ 165.46	 & \centering 162$^\circ$ \arraybackslash \\
\hline
\rule{0pt}{2.3ex} \centering \textbf{WCRK} \\ (0.44) & \centering 34.9$^\circ$ & \centering -85.7$^\circ$ & \centering 5 & \rule{0pt}{2.3ex} \centering 72.45 \\ $\scriptscriptstyle\pm$ 118.60 & \rule{0pt}{2.3ex} \centering 78.74 \\$\scriptscriptstyle\pm$ 123.81 & \centering 47$^\circ$ \arraybackslash \\
\hline
\rule{0pt}{2.3ex} \centering \textbf{\textit{BULL}} \\ (0.41) & \centering 36.1$^\circ$ & \centering -84.0$^\circ$ & \centering 3 & \rule{0pt}{2.3ex} \centering -59.84 \\ $\scriptscriptstyle\pm$ 113.50 & \rule{0pt}{2.3ex} \centering -79.15\\ $\scriptscriptstyle\pm$ 110.04 & \centering 233$^\circ$ \arraybackslash \\
\hline
\rule{0pt}{2.3ex} \centering \textbf{SHEL} \\ (0.26) & \centering 35.4$^\circ$ & \centering -89.8$^\circ$ & \centering 4 & \rule{0pt}{2.3ex} \centering -62.97 \\ $\scriptscriptstyle\pm$ 141.23 & \rule{0pt}{2.3ex} \centering 3.73 \\ $\scriptscriptstyle\pm$ 177.28 & \centering 177$^\circ$ \arraybackslash \\
\hline
\rule{0pt}{2.3ex} \centering \textbf{RCCN} \\ (0.22) & \centering 35.1$^\circ$ & \centering -85.4$^\circ$ & \centering 3 & \rule{0pt}{2.3ex} \centering -42.37 \\ $\scriptscriptstyle\pm$ 67.48 & \rule{0pt}{2.3ex} \centering -31.70 \\ $\scriptscriptstyle\pm$ 73.77	 & \centering 217$^\circ$ \arraybackslash \\
\hline
\rule{0pt}{2.3ex} \centering \textbf{\textit{BRAD}} \\ (0.19) & \centering 35.1$^\circ$ & \centering -84.9$^\circ$ & \centering 5 & \rule{0pt}{2.3ex} \centering -44.54 \\ $\scriptscriptstyle\pm$ 62.52 & \rule{0pt}{2.3ex} \centering 8.95\\ $\scriptscriptstyle\pm$85.13 & \centering 169$^\circ$ \arraybackslash \\
\hline
\rule{0pt}{2.3ex} \centering \textbf{\textit{RFRD}} \\ (0.14) & \centering 35.8$^\circ$ & \centering -86.6$^\circ$ & \centering 3 & \rule{0pt}{2.3ex} \centering 0.67 \\ $\scriptscriptstyle\pm$ 132.55 & \rule{0pt}{2.3ex} \centering -33.63 \\ $\scriptscriptstyle\pm$ 144.25 & \centering 271$^\circ$ \arraybackslash \\
\hline
\rule{0pt}{2.3ex} \centering \textbf{\textit{EPNT}} \\ (0.12) & \centering 34.2$^\circ$ & \centering -86.8$^\circ$ & \centering 1 & \rule{0pt}{2.3ex} \centering -25.79 \\ $\scriptscriptstyle\pm$ 37.46 & \rule{0pt}{2.3ex} \centering -14.35\\ $\scriptscriptstyle\pm$ 52.88 & \centering 209$^\circ$ \arraybackslash \\
\hline
\rule{0pt}{2.3ex} \centering \textbf{\textit{SULL}} \\ (0.09) & \centering 36.4$^\circ$ & \centering -82.3$^\circ$ & \centering 4 & \rule{0pt}{2.3ex} \centering 21.22 \\ $\scriptscriptstyle\pm$ 82.64 & \rule{0pt}{2.3ex} \centering 0.17 \\ $\scriptscriptstyle\pm$ 85.25 & \centering 0$^\circ$ \arraybackslash \\
\hline
\rule{0pt}{2.3ex} \centering \textbf{\textit{SHVN}} \\ (0.01) & \centering 35.0$^\circ$ & \centering -90.1$^\circ$ & \centering 4 & \rule{0pt}{2.3ex} \centering 1.80 \\ $\scriptscriptstyle\pm$ 184.68 & \rule{0pt}{2.3ex} \centering -3.00 \\ $\scriptscriptstyle\pm$ 184.24 & \centering 301$^\circ$ \arraybackslash \\
\hline
\multicolumn{7}{r}{\scriptsize{$^{\mathrm{a}}$Italicised nodes display multi-modal directionality, with effective average indicated}}\\
\multicolumn{7}{r}{\scriptsize{$^{\mathrm{b}}$Number of events ($n$) with data available}}\\
\end{tabular}
\label{tab:results}
\end{table}

In general terms, larger network parameters relate to larger susceptibility. PARA is the most susceptible node at around 241 Akm/V, with a defined risk rank of 1.0. This is twice as much as the next highest node, MONT, with a relative risk rank of 0.5. The ratio of the average network parameter spread and total network scaling gives an indication of the certainty of the estimate and complexity of the local network. WEAK has the most certainty in its network parameter ensemble, followed by BULL and WCRK. Italicised cases in Table \ref{tab:results} indicate nodes that have multiple lines influencing GIC exposure. The network parameters choose the most efficient and representative of these contributions, but in the directionality ensemble multiple peaks are evident and a larger spread in the network parameters ensembles is expected. Typically, these nodes are at complex or interior parts of the network and the multiple paths allow GICs to dissipate to non-critical levels, minimising susceptibility as seen in the relative risk ranking.

\section{Discussion}\label{sec:discuss}

From Tables \ref{tab:events} and \ref{tab:events_cum}, it is evident that apart from general GIC activity, the global SYM-H index is not always representative of the peak or cumulative GIC in a local network. A more local E-field is more appropriate for fine scale characterisation, as can be seen in the Sep 2014 event where the largest peak GIC ranks as smallest in terms of peak SYM-H, but largest in peak $E_x$ across events. Since PARA is a north-south effective node and the most susceptible in the TVA network, the large $E_x$ produces the peak GIC. More directly linked to SYM-H is the general east-west E-field tendency (modulated by local ground conductivity) of the ring current drivers in both impulsive and cumulative proxies. This difference between E-field components is particularly apparent in CME storms, when the ring current is most affected, with differences during CIR storms small in comparison.

Taking into account the ring current driving at mid-latitudes, with its most probable east-west E-field, risk is increased if the effective directionality of a node is east-west. From Table \ref{tab:results}, in the TVA network only RFRD is east-west. RFRD is an interior node with only a short transmission line and as such is low risk. WCRK and RCCN both have significant NE and SW contributions from the same part of the network and appear to link to the stronger east-west driving E-field during the Sep 2014 event, even though their network parameters are smaller than other nodes less affected during this event.

\begin{figure}[htbp]
\centerline{\includegraphics[width=3.38in]{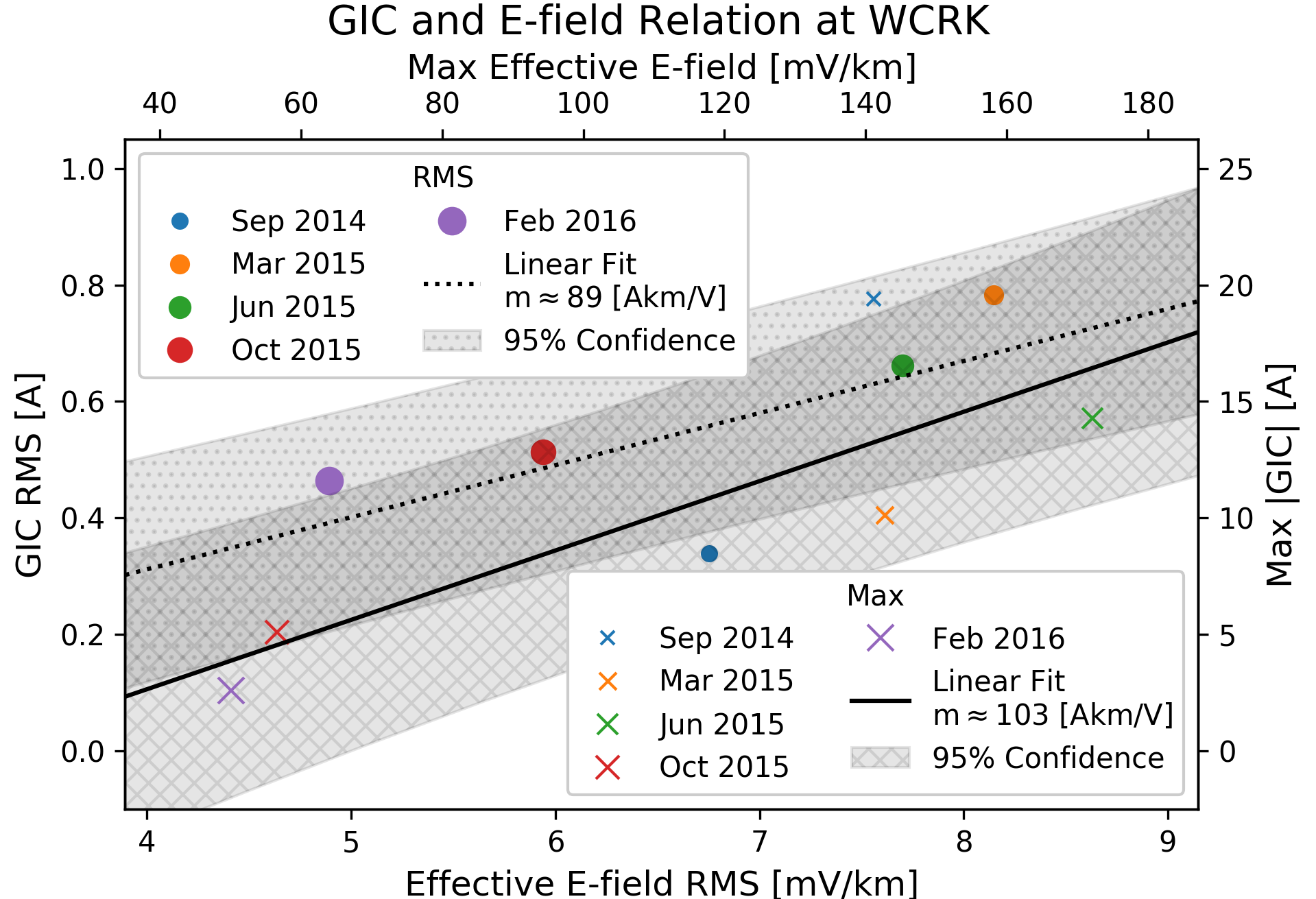}}
\caption{Linear trend between the peak and cumulative effective E-field and measured GIC at WCRK, which allows for extrapolation to other events.}
\label{fig:max_rel}
\end{figure}

Making use of the empirical network parameters that absorb errors in the geophysical modelling and network assumptions, the GIC response at a node can be related to more general parameters. Specifically, the network parameters allow for the effective direction to be determined and the scaled effective E-field contributions to be defined,
\begin{equation}
    E_{\text{eff}}=\cos{(\theta)}E_x+\sin{(\theta)}E_y. \label{eq:eff}
\end{equation}
Such a relation can be derived at a node for both peak and cumulative GIC and E-field exposure, as in Figure \ref{fig:max_rel}. 

The relation in both is linear, similar to the assumed linear network parameters that link E-field and GIC. Variance may arise from the maxima of the E-field components and GIC not occurring at the same time. At WCRK, Figure \ref{fig:max_rel} shows this trend is nevertheless consistent over all 5 events with the slopes of both cumulative and peak exposure relating to the absolute network scaling seen in Table \ref{tab:results}, i.e. $m\approx107$. The peak driving slope is larger than the RMS driving slope due to it only considering the largest contributions. Bulk analysis of more events will result in more accurate relations. In the case of a large deviation from the relation for an event, the most likely cause would be a network change such as line switching. A possible further cause may be a distinctly different structure of the geomagnetic storm. The Sep 2014 event is an example of such an outlier, with a particularly large impulsive peak and significantly smaller cumulative driving in comparison.

The relations between GIC and effective E-field being consistent, they can be used to extrapolate the local network exposure to existing extreme value analysis for the E-field in North America \cite{Lucas2019}. Similar analysis has been done using the time $dB/dt$ in New Zealand \cite{Rodger2017}. Using the 1-in-100 year E-field threshold of roughly 1 V/km estimated for the TVA region \cite{Lucas2019}, Table \ref{tab:results} can be interpreted as the resulting GIC in Amperes for the extreme E-field in the north and east directions respectively, with the peak exposure at the most susceptible node (PARA) being around 240 A. One step further is linking E-field to SYM-H \cite{Lotz2017} or its low-resolution twin, Dst \cite{Love2020}, and calibrating the local network exposure to a longer and more global dataset. More GIC event coverage is needed to validate such bulk relations locally.

Besides the extreme value exposure, given a typical solar cycle there are specific nodes that are susceptible in the TVA network. The most susceptible node is PARA, followed by WCRK, MONT, WEAK and RCCN that have elevated risk. These local edge nodes should be taken into account given mitigation efforts, with other nodes having negligible exposure. Similar cumulative damage risk is seen at PARA, WEAK, MONT and WCRK, which should inform maintenance scheduling. Any maintenance or mitigation efforts should take into account peak periods of geomagnetic storm activity, expected at solar maximum and the declining phase of the solar cycle \cite{Echer2011}. During these periods the associated GIC driving is able to initiate or accelerate accumulated damage.

Although this paper has focussed on mid-latitudes, where the bulk of power networks lie, a similar probabilistic network parameter ensemble approach can be applied to the more geomagnetically complicated high-latitudes.

\section{Conclusions}\label{sec:conc}
The probabilistic approach presented in this work is able to inform network-wide geomagnetic risk analysis without the need for in-depth network information or complex ground conductivity modelling. Network parameter ensembles are derived using limited measured GIC and B-field data and form distributions, rather than typical transformer-level or extreme value modelling that use single value network parameters in the engineering step. Given the FERC directive for utilities to collect measured GIC data \cite{FERC}, the approach employed is widely applicable. Nodal and network vulnerability can be identified and calibrated through a general E-field. The resulting calibration in the TVA network is extrapolated to extreme value E-fields and given a 1-in-100 year scenario, GICs of over 200 A at a single node and around 100 A at four others may be experienced. Using the empirical calibration of the engineering step, a probability distribution of GIC magnitude for an existing node can possibly be derived directly from a probability distribution of storm severity.

\section*{Acknowledgment}
The authors acknowledge the Tennessee Valley Authority for measured GIC data. Geomagnetic field data were collected at Fredericksburg. We thank the USGS for supporting the operation of this geomagnetic observatory and INTERMAGNET for promoting high standards of magnetic observatory practice (\url{www.intermagnet.org}). The SYM-H index is provided by NASA/GSFC Space Physics Data Facility's OMNIWeb service (\url{omniweb.gsfc.nasa.gov}). SSC onsets are part of the calculated SC index, made available by Observatori de l'Ebre, Spain, from data collected at magnetic observatories. We thank the involved national institute and ISGI (\url{isgi.unistra.fr}).


\end{document}